\documentclass{appolb}
\usepackage{graphicx}

\begin{document}
\def\Pom{\mathrm{I\!P}}

\title{Exclusive diffractive dijets at HERA and EIC using GTMDs%
\thanks{Presented by A.S. at Diffraction and Low-x 2024 workshop, 
Trabia, Italy}%
}
\author{Antoni Szczurek
\address{Institute of Nuclear Physics PAN, PL-34-342 Krak\'ow, Poland 
and Rzesz\'ow Univeristy, PL-35-310 Rzesz\'ow, Poland}
\\[3mm]
Barbara Linek
\address{College of Mathematics and Natural Sciences,
University of Rzesz\'ow, PL-35-310 Rzesz\'ow, Poland}
}
\maketitle
\begin{abstract}
We calculate differential distributions for diffractive production 
of dijets in $ep\rightarrow e^{'}p\,jet\,jet$ reaction using 
off diagonal unintegrated gluon distributions, often called GTMDs 
for brevity. Different models are used. We focus on 
the contribution to exclusive $q\bar{q}$ dijets.

The results of our calculations are compared with the H1 and ZEUS data. 
Except of one GTMD, our results are below the HERA data points. 
This is in contrast with recent results where the normalization 
was adjusted to some selected distributions and no agreement 
with other observables was checked. We conclude that the calculated 
cross sections are only a small part of the measured ones 
which probably contain also processes with pomeron remnant, 
reggeon exchange, etc.

We present also azimuthal correlations between the sum and 
the difference of dijet transverse momenta. The cuts on transverse 
momenta of jets generate azimuthal correlations (in this angle) which
can be easily misinterpreted as due to so-called elliptic GTMD.
\end{abstract}
 
\section{Introduction}

This work focuses on exclusive, diffractive production of dijets in 
the $ep \to ejjp$ reaction, where the final-state proton remains 
in its ground state. This presentation is based on our recent
publication \cite{our_dijets}. The processes discussed there were 
measured by the H1 \cite{H1:2012} and ZEUS \cite{ZEUS:2016} 
collaborations. We use a formalism derived 
from the color dipole approach but the dipole amplitude information 
from impact parameter space is mapped to off-forward transverse 
momentum-dependent gluon distributions (GTMDs). For reviews 
linking this to the gluon Wigner function, see 
\cite{Pasechnik:2023mdd}. At large jet transverse momenta, the forward 
diffractive amplitude directly probes the unintegrated gluon 
distribution of the target \cite{Nikolaev:1994cd, Nikolaev:2000sh}. 
While this approach is suited for the small-$x$ limit, longitudinal 
momentum transfer and skewedness are handled in a collinear 
factorization framework using generalized parton distributions, 
as in \cite{Braun:2005rg}. This work includes also $q\bar{q}$ 
exchanges in the $t$-channel, relevant for smaller rapidity gaps.

In \cite{Linek:2023kga}, we applied various GTMD models to 
the $pA \to c\bar{c}pA$ process, although no data is available yet 
for this reaction due to several challenges of relevant measurements. 
Here, we apply the same formalism to $e p \to j j p$ in order to 
confront our results with the H1 and ZEUS data, comparing results of 
different GTMD models.

Recent theoretical calculations on diffractive dijet production, 
using either the color dipole or GTMD approaches can be found in 
\cite{PhysRevD.100.074020, Hagiwara:2016kam, Hagiwara:2017fye,
  ReinkePelicer:2018gyh, Boer:2021upt, Boer:2023mip}. 
Some of these works focus on photoproduction of dijets 
or production of heavy quarks. 
Our study has some overlap with \cite{PhysRevD.100.074020}, 
which uses the Golec-Biernat--W\"usthoff parametrization 
\cite{Golec-Biernat:1998zce} for the dipole amplitude. 
For the corresponding gluon distribution, our results agree 
with the other results. We employ also the GTMDs proposed and 
fitted in \cite{Boer:2021upt, Boer:2023mip}.
However, our conclusions differ from those works.

\section{Sketch of the formalism}

To calculate the cross section for $ep\rightarrow ep\,q\bar{q}$ both 
the transverse $\sigma_{T}$ and longitudinal $\sigma_{L}$ cross 
sections have to be included:
\begin{eqnarray}
\frac{d\sigma^{ep}}{dy dQ^{2} d\xi} &=&\frac{\alpha_{em}}{\pi y Q^{2}} \Big[ \Big(1-y+\frac{y^{2}}{2} \Big) \frac{d \sigma_{T}^{\gamma^{*}p}}{d \xi}+(1-y) \frac{d\sigma_{L}^{\gamma^{*}p}}{d\xi}  \Big],
\end{eqnarray}
where $d \xi = dz d^2\vec P_\perp d^2 \vec \Delta_\perp$, while the 
interferences between photon polarizations are neglected as they 
vanish when averaging over the angle between the electron scattering 
and the hadronic planes.

 \begin{figure}
  \centering
  \includegraphics[width=.242\textwidth]{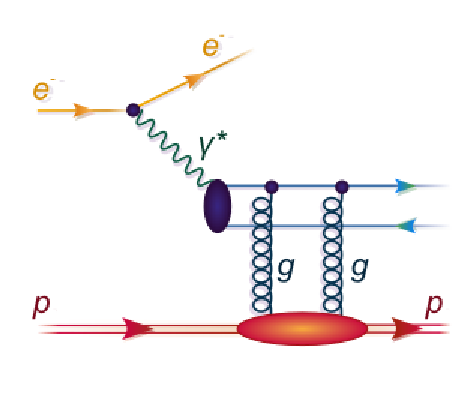}
  \includegraphics[width=.242\textwidth]{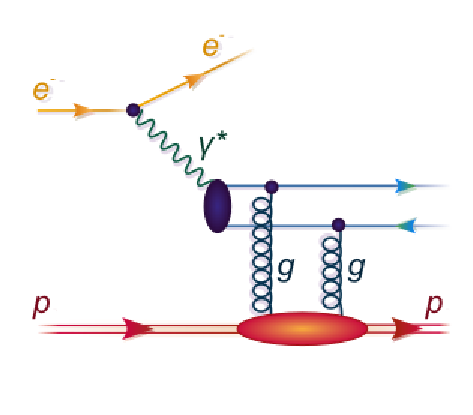}
  \includegraphics[width=.242\textwidth]{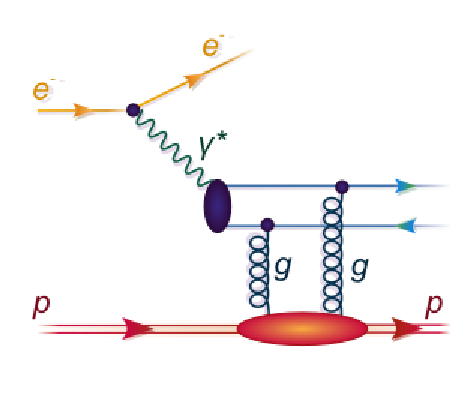}
  \includegraphics[width=.242\textwidth]{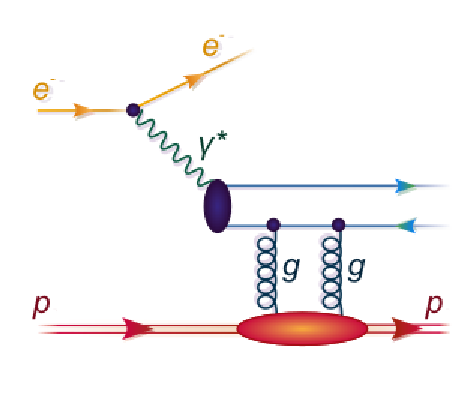}  
  \caption{Four Feynman diagrams for the diffractive production 
of dijets in electron-proton collisions.}
  \label{fig:diagrams}
\end{figure}
For all four mechanisms shown in Fig.\ref{fig:diagrams}, the $\gamma^* p \to q \bar q p$ cross sections for transverse and longitudinal photons are given by:
\begin{eqnarray}
\frac{d\sigma^{\gamma^{*}p}_{T}}{dzd^{2}\vec P_{\perp}d^{2} \vec \Delta_{\perp}}&=&2N_{c}\alpha_{em}\sum_{f}e_{f}^{2}\int d^{2}\vec k_{\perp}\int d^{2}\vec k^{~'}_{\perp}T \big(Y,\vec k_{\perp},\vec \Delta_{\perp}\big) T\big(Y,\vec k^{~'}_{\perp},\vec \Delta_{\perp} \big) \nonumber\\ 
 &\times& \Bigg\{ \Big(z^2 + (1-z)^2 \Big)\left[\frac{(\vec P_{\perp}-\vec k_{\perp})}{(\vec P_{\perp}-\vec k_{\perp})^{2} +\epsilon^2}-\frac{\vec P_{\perp}}{P^{2}_{\perp}+\epsilon^2}\right] \nonumber \\
 &\cdot& \left[\frac{(\vec P_{\perp}-\vec k^{~'}_{\perp})}{(\vec P_{\perp}-\vec k^{~'}_{\perp})^{2} + \epsilon^2}-\frac{\vec P_{\perp}}{P^{2}_{\perp}+\epsilon^2}\right] \nonumber \\
 &+& m_f^2 \left[\frac{1}{(\vec P_{\perp}-\vec k_{\perp})^{2}
 +\epsilon^2}-\frac{1}{P^{2}_{\perp}+\epsilon^2}\right] \nonumber \\
 &\cdot& \left[\frac{1}{(\vec P_{\perp}-\vec k^{~'}_{\perp})^{2}+\epsilon^2}-\frac{1}{P^{2}_{\perp}+\epsilon^2}\right] \Bigg\},
\end{eqnarray}
\begin{eqnarray}
\frac{d\sigma^{\gamma^{*}p}_{L}}{dzd^{2}\vec P_{\perp}d^{2} \vec\Delta_{\perp}}&=&2N_{c}\alpha_{em} 4Q^2 z^2 (1-z)^2 \nonumber \\
&\times& \sum_{f}e_{f}^{2}\int d^{2}\vec k_{\perp}\int d^{2}\vec k^{~'}_{\perp}T \big(Y,\vec k_{\perp},\vec \Delta_{\perp}\big) T\big(Y,\vec k^{~'}_{\perp},\vec \Delta_{\perp} \big) \nonumber\\ 
 &\times& \left[\frac{1}{(\vec P_{\perp}-\vec k_{\perp})^{2} + \epsilon^2}-\frac{1}{P^{2}_{\perp}+\epsilon^2 }\right] \nonumber \\
 &\cdot& \left[\frac{1}{(\vec P_{\perp}-\vec k^{~'}_{\perp})^{2} + \epsilon^2}-\frac{1}{P^{2}_{\perp}+\epsilon^2}\right],
\end{eqnarray}
with $\epsilon^{2}=z(1-z)Q^{2}+m_f^2$ and the generalized transverse
momentum distribution (GTMD) of gluons in the proton target are 
expressed as a Fourier transform of the diffraction amplitude in 
momentum space 
(see \cite{Hagiwara:2016kam, Hagiwara:2017fye, ReinkePelicer:2018gyh, Pasechnik:2023mdd}):
\begin{eqnarray}
    T(Y,\vec k_\perp, \vec \Delta_\perp) = \int \frac{d^2 \vec b_\perp}{(2 \pi)^2} \frac{d^2 \vec r_\perp}{(2 \pi)^2}  
    e^{-i \vec \Delta_\perp \cdot \vec b_\perp} e^{-i \vec k_\perp \cdot \vec r_\perp} \, N(Y,\vec r_\perp, \vec b_\perp) \, e^{-  \varepsilon r_\perp^2}\, .
     \label{eq:reg}
\end{eqnarray}
The used normalization is consistent with that in
Ref.~\cite{ReinkePelicer:2018gyh} and 
the regularization parameter $\varepsilon = (0.5 \, \rm{fm})^{-2}$ 
is used in the calculation. 
We also analyzed special correlations in azimuthal angle between 
the sum and difference of transverse momenta of jets:
\begin{eqnarray}
\cos \phi_{\vec P_\perp \vec \Delta_\perp}  = \frac{\vec P_\perp \cdot \vec \Delta_\perp}{P_\perp \Delta_\perp} \, ,
\end{eqnarray}
where 
\begin{eqnarray}
    \vec P_\perp = \frac{1}{2}( \vec p_{\perp 1} - \vec p_{\perp 2} )\, , \qquad 
    \vec \Delta_\perp = \vec p_{\perp 1} + \vec p_{\perp 2} \, . 
\end{eqnarray}

In \cite{our_dijets} we considered six different models for generalized 
transverse momentum distributions (GTMDs). 
Two of these are parameterizations of 
off-forward gluon density matrices based on diagonal unintegrated 
gluon distributions: 
Golec-Biernat–W\"usthoff (GBW) model \cite{Golec-Biernat:1998zce} and 
Moriggi-Paccini-Machado (MPM) model \cite{Moriggi:2020zbv}. 
Both use a diffractive slope of $B = 4$ $\rm GeV^{-2}$:
\begin{eqnarray}
     f\Big(Y,\frac{\vec \Delta_\perp}{2} + \vec k_\perp, \frac{\vec \Delta_\perp}{2} - \vec k_\perp\Big) = \frac{\alpha_s}{4 \pi N_c} \, \frac{{\cal F}(x_\Pom, \vec k_\perp, -\vec k_\perp)}{k_\perp^4} \, \exp\Big[ - \frac{1}{2} B \vec \Delta^2 \Big] \, . 
     \label{eq:model_f}
\end{eqnarray}
The other four distributions are derived from the Fourier transform of 
the dipole amplitude described by equation (\ref{eq:reg}).

We use also the bSat model of Kowalski and Teaney 
\cite{Kowalski:2003hm} (KT model), as well as three models based 
on the McLerran-Venugopalan (MV) approach \cite{McLerran:1993ka}. 
These include the Iancu-Rezaeian model (MV-IR) \cite{Iancu:2017fzn}, 
the Boer-Setyadi 2021 model (MV-BS 2021) \cite{Boer:2021upt}, 
and the Boer-Setyadi 2023 model (MV-BS 2023) \cite{Boer:2023mip}, 
which were fitted to the H1 experimental data. In addition, we modified the MV-IR model using $\lambda = 0.277$:
\begin{eqnarray}
    T_{\rm MV-IR}^{\rm mod}(Y,\vec k_\perp, \vec \Delta_\perp) = T_{\rm
      MV-IR}(\vec k_\perp, \vec \Delta_\perp) \, e^{\lambda Y}, \quad \,
    Y = \ln\Big[\frac{0.01}{x_\Pom}\Big] \,.
\end{eqnarray}
To adopt the MV-BS 2021 to describe the H1 data \cite{H1:2012} we added 
according to ~\cite{Boer:2021upt} $\chi=1.25$ in 
the expression:
\begin{eqnarray}
N_0(r_\perp, b_\perp)=-\frac{1}{4}r_{\perp}^{2}\chi Q_{s}^{2}(b_{\perp})\ln \Big[\frac{1}{r_{\perp}^{2}\lambda^{2}} +e \Big], \,  Q_{s}^{2}(b_{\perp})=\frac{4\pi\alpha_{s}C_{F}}{N_{c}} \exp \Big[ \frac{-b_{\perp}^{2}}{2R_{p}^{2}} \Big].
\end{eqnarray}
For MV-BS 2023 the 
$\chi(x_{Bj})=\bar{\chi}\Big(\frac{x_{0}}{x_{Bj}}\Big)^{\lambda_{\chi}},$
 where $\bar{\chi}=1.5,\,x_{0}=0.0001$ and $\lambda_{\chi}=0.29$ 
are used according to ~\cite{Boer:2023mip}.

\section{Selected results}

\begin{figure}
\centering
\includegraphics[width=.449\textwidth]{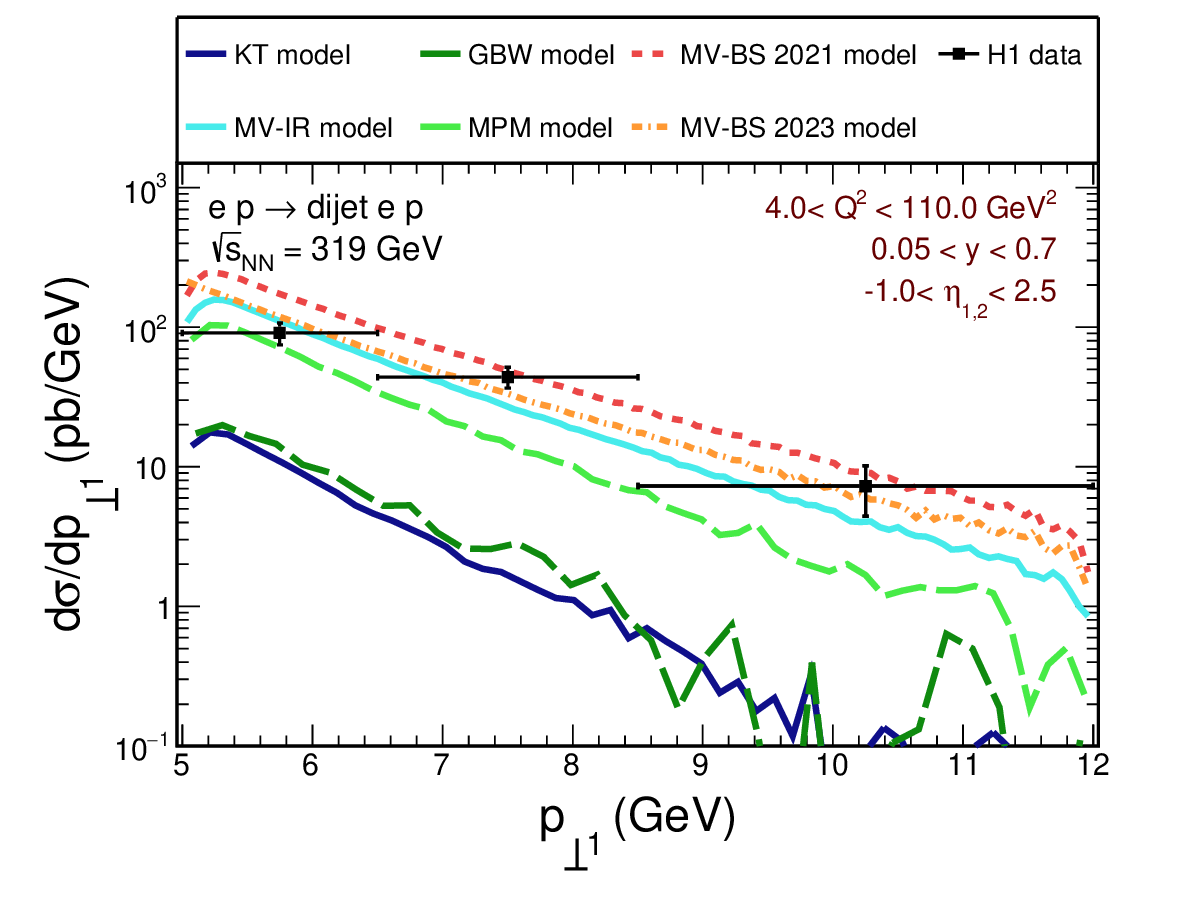}
\includegraphics[width=.449\textwidth]{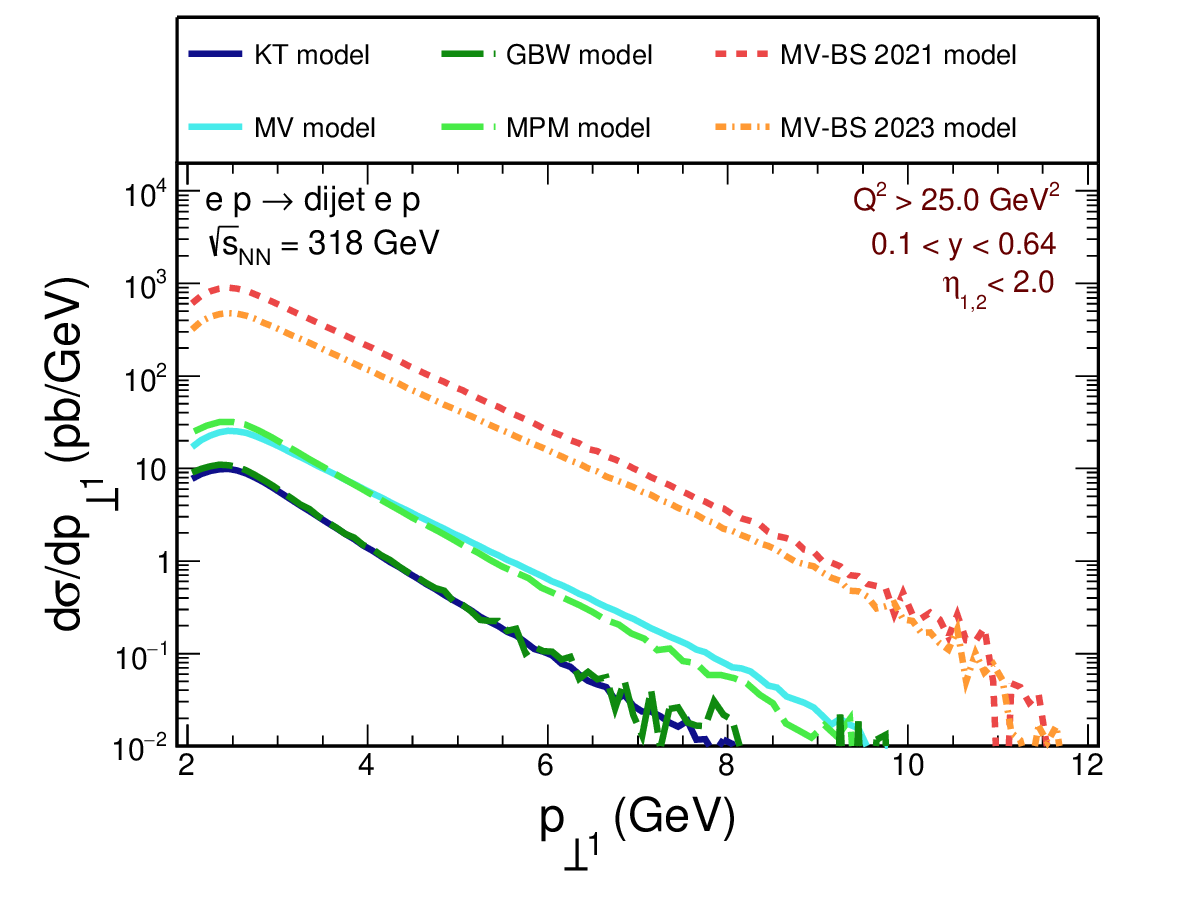}
\caption{Distribution of the cross-section for the diffractive light-quark 
dijet production in jet transverse momentum for H1 (left) and ZEUS
(right) kinematics for different GTMDs.}
\label{fig:epjj_dsig_dpt1}
\end{figure}

\begin{figure}
\centering
\includegraphics[width=.449\textwidth]{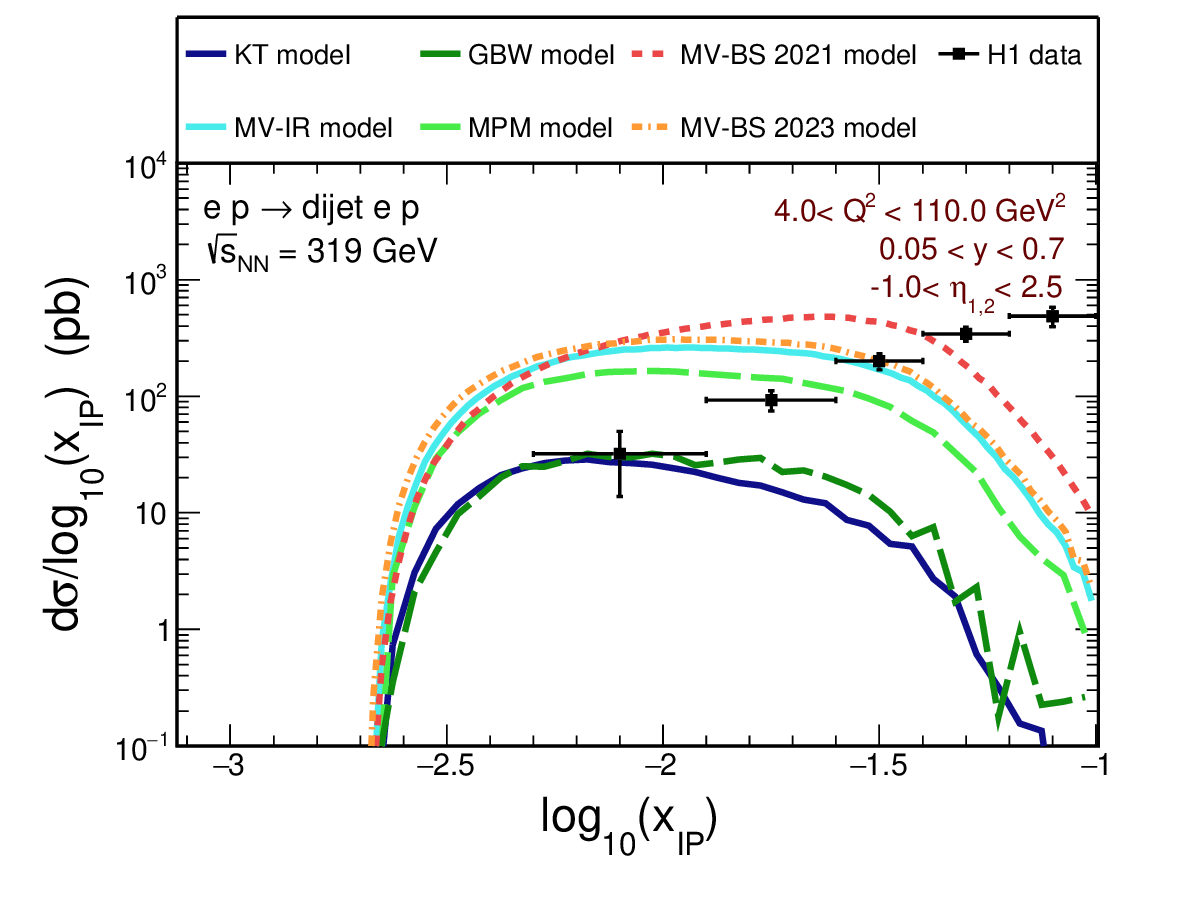}
\includegraphics[width=.449\textwidth]{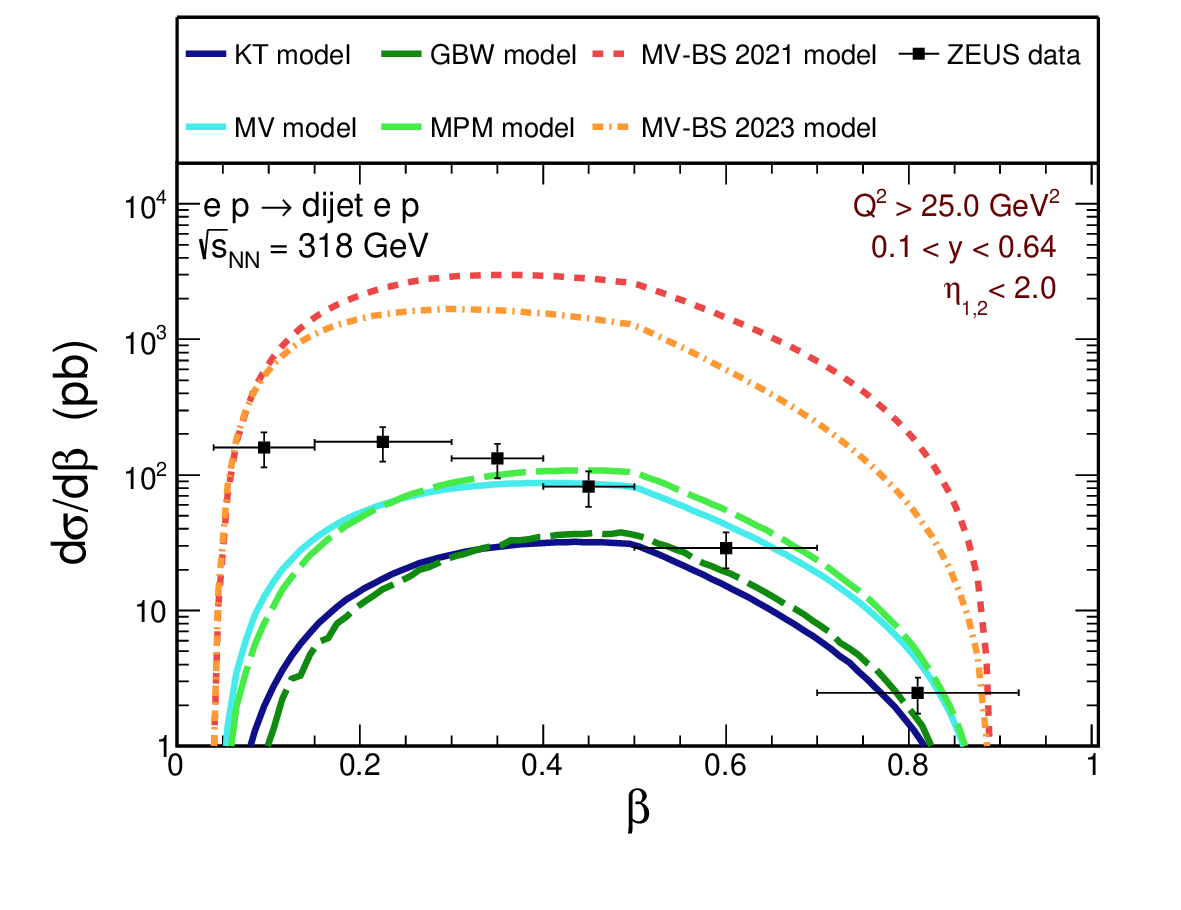}
\caption{Distribution of the cross-section for the diffractive 
light-quark dijet production in $x_{\Pom}$ and $\beta$ for H1 and 
ZEUS kinematic for different GTMDs.}
\label{fig:epjj_dsig_dlog10xpbeta}
\end{figure}

Our calculations were divided into two areas according to the kinematics
of the H1 and ZEUS collaborations. We first show the distributions 
in the transverse momentum of the jet shown in
Fig.~\ref{fig:epjj_dsig_dpt1}. 
The MV-BS 2021 and MV-BS 2023 give similar results to the MV-IR and MPM 
models and describe the data quite well, while the KT and GBW
distributions are lower by an order of magnitude than the experimental
data. Both the MV-BS results for the ZEUS kinematics differ by almost 
two orders of magnitude from the results of other GTMDs, 
however, the shapes of all distributions are similar. 
We also generated distributions in $x_{\Pom}$ and $\beta$ shown 
in Fig. \ref{fig:epjj_dsig_dlog10xpbeta}, where the differences between 
all models are visible. In the case of the dependence on 
$x_{\Pom}$, the data are overestimated by all GTMD models except 
of those based on KT and GBW UGDFs, see Eq.(\ref{eq:model_f}). 
This may be related to the fact that correct description 
of all experimental data requires considering not only the dipole
approach but also the contribution of $q\bar{q}$ exchanges,
see e.g. Ref.~\cite{Luszczak:2014cxa}. 

The distributions in $\beta$ also show inconsistencies with the
experimental data 
for the MV-BS models that were fitted to the H1 experiment. 
In contrast, the other models give results that are below experimental
data for small $\beta$, 
However, this area can be sensitive to the $q\bar{q}g$ 
three-parton contributions.

\begin{figure}
\centering
\includegraphics[width=.449\textwidth]{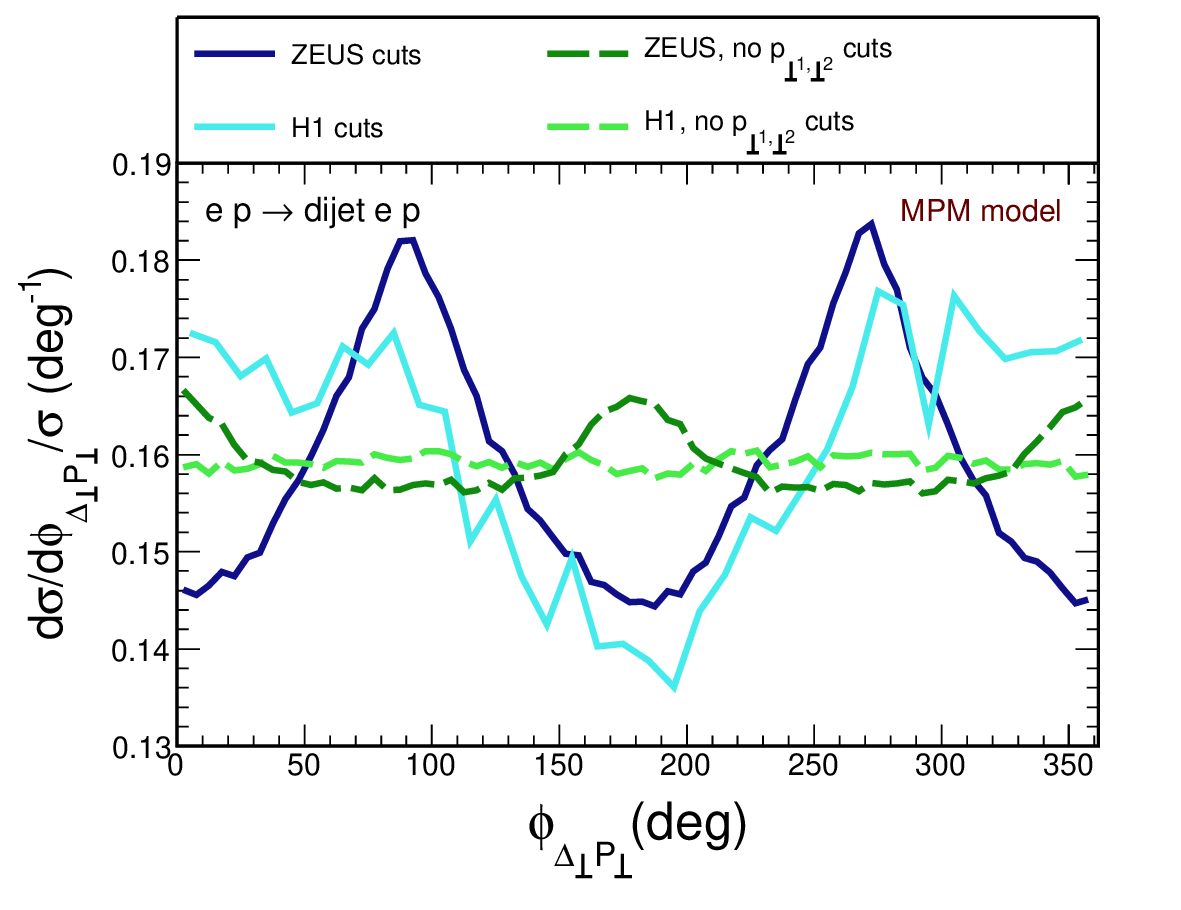}
\includegraphics[width=.449\textwidth]{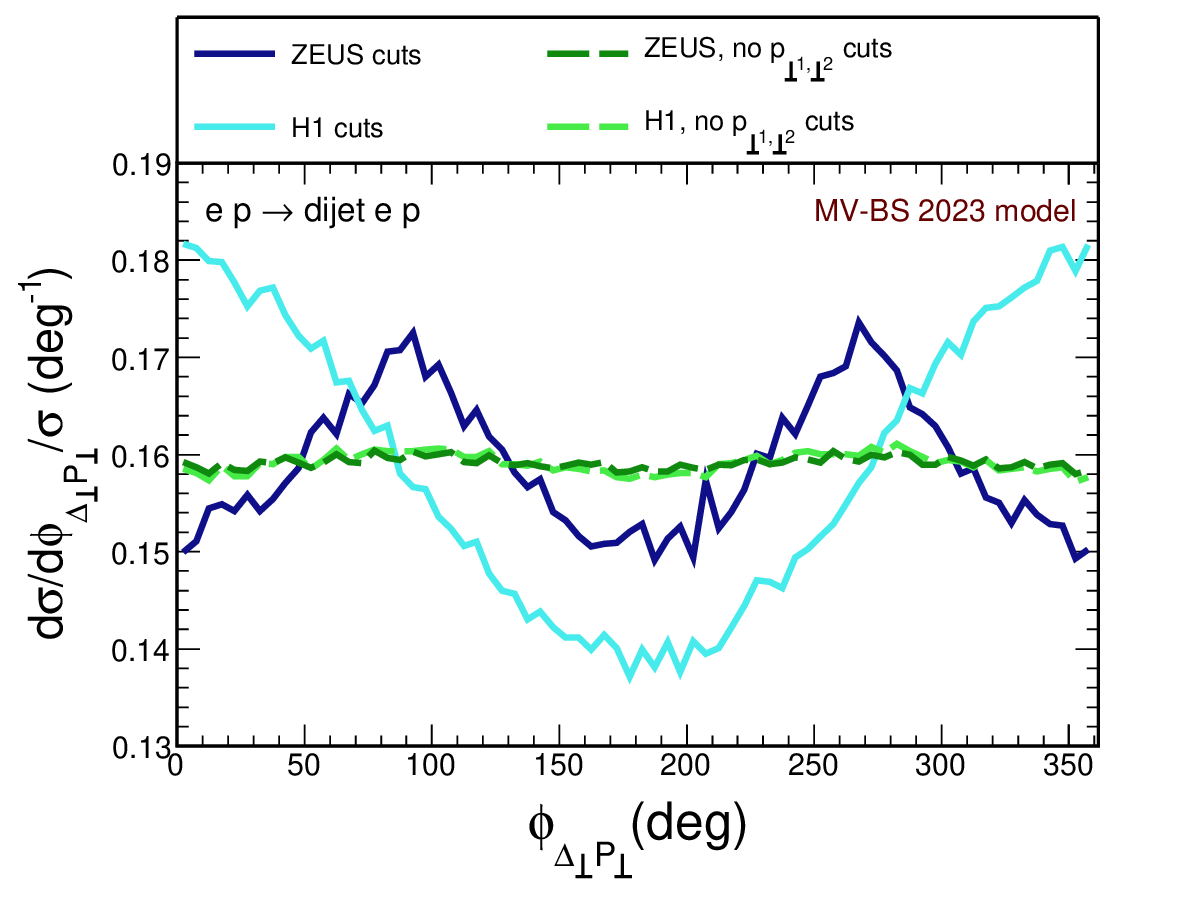}
\caption{Distribution of the cross-section for the diffractive 
light-quark dijet production in the energy of the photon-proton 
system  (left) and 
azimuthal angle $\phi$ between $\vec{P}_{\perp}$ and 
$\vec{\Delta}_{\perp}$ (right) for H1 and ZEUS kinematic for different
GTMDs. The reader is asked to notice the normalization.}
\label{fig:epjj_dsig_dphi}
\end{figure}

In Fig.~\ref{fig:epjj_dsig_dphi} we show the distribution of the
azimuthal angle between the sum and difference of the jets' transverse
momenta. 
We predict that the straight horizontal line corresponds to the case
without cuts on the transverse momentum of the jets, while the angular 
correlations can be seen for the situation in which such cuts 
are included. 
We do not exclude the possibility that the additional azimuthal
correlation may be due to elliptical gluon distributions, which 
were not taken into account in \cite{our_dijets}. 

\section{Conclusions}

We have discussed dijet production in the $ep\rightarrow epjj$ process. 
The corresponding differential distributions have been calculated 
using various gluon GTMD (generalized transverse momentum 
dependent gluon distributions) from the literature. We have 
calculated the distributions in various kinematic variables 
by referring to H1 and ZEUS data. 
The MV-BS, MPM, and MV-IR GTMD distributions describe some of 
the observables quite well but do not describe the distributions 
in $x_\Pom$ and $\beta$. Some of the other GTMD distributions 
are consistent with the H1 and ZEUS data. 
In our opinion the most realistic gluon 
distributions are those based on KT and GBW, which give 
rather small contribution for the H1 kinematics, 
and a sizable contribution at $\beta > 0.5$ for the ZEUS cuts. 
We conclude that the considered gluonic mechanism is not sufficient.
Therefore we plan to continue the topic.

We have also calculated correlations in azimuthal angles between 
the sum and difference of the jet's transverse momenta. Since our 
GTMDs do not have an elliptical part, these correlations are solely 
the result of experimental cuts. 

\vspace{0.5cm}

{\bf Acknowledgements}

A.S. is indebted to Marta {\L}uszczak and Wolfgang Sch\"afer
for collaboration on the issues presented here.


\end{document}